\documentclass[11pt,letterpaper,abstract=true]{scrartcl}
\usepackage[margin=1in]{geometry}
\usepackage[T1]{fontenc}
\usepackage{lmodern}
\usepackage{microtype}
\usepackage[UKenglish]{babel}
\usepackage{csquotes}
\usepackage{mathtools}
\usepackage{amssymb,amsthm}
\usepackage{thm-restate}
\usepackage{fontawesome5}
\usepackage{graphicx}
\usepackage{booktabs}
\usepackage{xcolor}
\usepackage{bm}
\usepackage{tabularx}
\usepackage{etoolbox}
\usepackage{enumitem,needspace}
\usepackage{algorithm}
\usepackage{algpseudocode}
\usepackage{tikz}
\usepackage{xurl}
\usepackage[hidelinks,unicode]{hyperref}
\usepackage[capitalise,noabbrev]{cleveref}
\usepackage{todonotes}
\usepackage[
  backend=biber,
  style=alphabetic,
  minalphanames=3,
  maxalphanames=4,
  maxbibnames=99,
  doi=true,
  giveninits=true
]{biblatex}
\hypersetup{
  colorlinks=true,
  linkcolor=blue!40!black,
  citecolor=blue!40!black,
  urlcolor=blue!40!black,
  pdftitle={A 2.8027-Approximation for Moldable Precedence Scheduling},
  pdfsubject={Moldable scheduling, precedence constraints, approximation algorithms},
  pdfkeywords={moldable scheduling, linear programming, threshold rounding, list scheduling}
}
\numberwithin{equation}{section}
\theoremstyle{plain}
\newtheorem{theorem}{Theorem}
\newtheorem{lemma}[theorem]{Lemma}

\newtheorem{claim}{Claim}
\crefname{claim}{claim}{claims}
\Crefname{claim}{Claim}{Claims}
\theoremstyle{definition}
\newtheorem{definition}[theorem]{Definition}
\theoremstyle{remark}

\newtheorem*{claim*}{Claim}
\newcommand{\OPT}{\operatorname{OPT}}
\newcommand{\Cmax}{C_{\max}}

\setkomafont{disposition}{\normalfont\bfseries}
\setkomafont{title}{\normalfont\LARGE\bfseries}
\title{Below Three: A 2.8027-Approximation for Moldable Precedence Scheduling}

\author{%
    Klaus Jansen\thanks{Kiel University, kj@informatik.uni-kiel.de, \url{https://orcid.org/0000-0001-8358-6796}}
    \and
    Felix Ohnesorge\thanks{Kiel University, foh@informatik.uni-kiel.de, \url{https://orcid.org/0009-0003-8023-3380}}
}

\date{}
\AfterEndEnvironment{abstract}{%
  \par\smallskip
  {\small\noindent\textbf{Keywords.} Moldable scheduling, precedence
  constraints, approximation algorithms, linear programming.
  \par
  \small\noindent\textbf{Funding.} Deutsche Forschungsgemeinschaft (DFG, German Research Foundation) -- Project number 453769249.}%
}
\begin{document}
\maketitle

\begin{abstract}
      We study makespan minimisation for nonpreemptive moldable jobs with
      arbitrary precedence constraints on identical processors. Each job
      runs on a chosen number of processors that remains fixed throughout
      its execution. Allocating more processors to a job can shorten its
      execution, but may increase total work and delay other jobs.
      We assume nonincreasing processing times, nondecreasing work, and a
      convex duration--work relation after dominated allocations are
      removed. This model includes jobs with concave speedup, for which
      Jansen and Zhang obtained a $3.291919\ldots$-approximation.
      Independently and concurrently with our work, Chen obtained a
      $3.016208\ldots$-approximation for the full convex model and a
      $2.973912\ldots$-approximation for concave speedup.
      We give a deterministic polynomial-time $2.8027$-approximation for
      the full convex model and every number of processors.
      Our result breaks the factor-three barrier for this model and also
      improves the approximation guarantee for concave speedup.
\end{abstract}

\section{Introduction}\label{pub-introduction}

Parallel applications combine two forms of concurrency: several jobs may run at the same time, and a single job may use several processors.
A scheduler must decide how to divide the available processors between these alternatives.
Allocating more processors to a job can shorten a chain of dependent jobs, but imperfect speedup may increase the total processor time consumed.
With precedence constraints, a locally attractive allocation can therefore delay other jobs and increase the overall completion time.
The central difficulty is to choose allocations that balance precedence-path lengths and total work, and then turn those allocations into a feasible schedule.

We study nonpreemptive \emph{moldable} scheduling in the following model.

\begin{definition}[Moldable Scheduling with Precedence Constraints]
      An instance consists of a directed acyclic graph $G=(J,E)$, $m$ identical processors, and positive rational processing times $t_j(p)$ for every job $j\in J$ and every integer $1\le p\le m$.
      A schedule chooses an allocation $p_j$ and a start time $S_j\ge0$ for each job.
      Job $j$ occupies $p_j$ processors throughout $[S_j,S_j+t_j(p_j))$.
      At every time, the total number of occupied processors is at most $m$, and $S_j\ge S_i+t_i(p_i)$ for every $(i,j)\in E$.
      The objective is to minimise the makespan $\Cmax=\max_j\{S_j+t_j(p_j)\}$.
\end{definition}

Allocations remain fixed during execution, jobs cannot be preempted, and processor indices need not be contiguous.
The input explicitly contains the processing-time tables.
Earlier papers also use the term nonpreemptive \emph{malleable} for this fixed-allocation model~\cite{jansen2012}; we use \emph{moldable} to distinguish it from models in which allocations may change during execution.

We assume that processing time is nonincreasing and that work, the product of allocation and processing time, is nondecreasing.
Thus, for every job $j$ and $1\le p<m$,
\[
      t_j(p+1)\le t_j(p),\qquad
      p\,t_j(p)\le(p+1)t_j(p+1).
\]
After discarding allocations that have the same duration as a smaller allocation, we additionally assume that the piecewise-linear curve through all remaining duration--work points is convex.
This model includes the concave-speedup setting of Jansen and Zhang~\cite{jansen2012} and permits duration plateaus without imposing a particular speedup
formula.
\Cref{def:pub-frontier} formalises this convexity condition.

\paragraph{Our Result.}
We give an approximation guarantee below $2.803$ for every number of processors, including instances with arbitrarily many jobs and arbitrary precedence constraints.

\begin{theorem}\label{pub-main-theorem}
      For moldable jobs with the preceding monotonicity and convexity properties and arbitrary precedence constraints, there is a deterministic polynomial-time $\rho$-approximation for every number of processors, where
      \begin{equation}\label{eq:exact-ratio}
            \rho=\frac{2802671}{10^6}=2.802671<2.8027.
      \end{equation}
\end{theorem}

The algorithm solves one linear program and constructs a fixed finite family of schedules from the solution to this linear program.
Each schedule is obtained by rounding the fractional solution using specific parameters and applying list scheduling.
The algorithm returns the shortest resulting schedule.
The rounding parameters are specified by constants independent of the instance, and the algorithm is deterministic.

The separation between choosing allocations and scheduling the resulting parallel jobs has a long history.
Turek, Wolf, and Yu~\cite{turek1992} used this framework for independent parallelisable tasks.
With precedence constraints, the allocation problem also resembles the discrete time-cost tradeoff problem: selecting a faster execution mode increases its cost, while the duration of every path must be controlled.
Skutella's bicriteria approximation algorithms~\cite{skutella1998} address this tradeoff.
In moldable scheduling, the work plays the role of cost, and the chosen modes must additionally be scheduled subject to the processors' instantaneous capacity.
List scheduling analyses connect these two stages through path-length and workload bounds, following the classical reasoning of Graham~\cite{graham1966}.

Under the two monotonicity assumptions alone, Lep\`ere, Trystram, and Woeginger~\cite{lepere2002} obtained a $(3+\sqrt5)$-approximation for arbitrary precedence constraints.
Jansen and Zhang~\cite{jansen2006} improved this factor to $4.730598\ldots$ by refining the relaxation and rounding analysis.
Chen~\cite{chen2018} subsequently studied an iterative LP method for this general monotone model, reporting a factor of $4.4841$ after two iterations and factors approaching $2+\sqrt2$ as the number of iterations increases.
These results do not assume the convex duration-work frontiers required here.

Additional structure in the processing-time tables supports stronger bounds.
Jansen and Zhang~\cite{jansen2012} showed that concave speedup implies convexity of the duration-work relation and obtained a $3.291919\ldots$-approximation by LP rounding, processor capping, and list scheduling.
Their rounding rule uses the interpolation weight between adjacent durations.
Chen and Chu~\cite{chenchu2013} considered convex work as a function of duration and proved a $(2+\sqrt2)$-approximation.
Their paper also reported a factor of $2.9549$ for strictly decreasing processing times; Chen's recent preprint~\cite{chen2026} identifies an invalid elimination step in that argument and corrects that claim.

\paragraph{Independent and Concurrent Work.}
Chen's arXiv preprint~\cite{chen2026} studies the same convex-frontier model and uses similar ideas: a rounding decision based on the slower duration relative to the fractional duration, followed by processor capping and list scheduling.
Our results and Chen's were obtained independently and concurrently.
Chen proves an optimal pointwise time-work tradeoff for this rounding rule and obtains a $13(30+\sqrt{493})/225=3.016208\ldots$-approximation for the general convex-frontier model and a $2.973912\ldots$-approximation for concave speedup.
Our improvement comes from averaging work and schedule bounds across a family of threshold-cap pairs and from excluding path work from the waiting-time charge.
The factor in~\eqref{eq:exact-ratio} is smaller than both of these guarantees and applies to the full model considered here.

\paragraph{Organisation.}
\Cref{pub-main-section} presents the LP relaxation, followed by rounding and list scheduling with parameters left unspecified.
\Cref{sec:analysis} gives the parameterised analysis, and \Cref{sec:parameters} explains the optimisation, the numerical search, and an exactly certified choice.
\Cref{sec:final-guarantee} completes the approximation and running-time analysis.
The main algorithm and its parameter selection concern $m\ge6$; Appendix~\ref{app:small} handles $m\le5$ with a stronger bound.
Algebraic proofs and the details of the exact certificate are deferred to the appendix.
The basic relaxation and list-scheduling lemmas remain valid for every positive $m$.

\section{The Algorithm}\label{pub-main-section}

Write $[k]=\{1,\ldots,k\}$, and let $\OPT$ denote the minimum makespan.
Polynomial time refers to the binary encoding length of the explicit tables and the precedence graph.
Paths are nonempty; a single job is a path.
We exclude empty instances, whose optimum is zero.

We describe the algorithm for $m\ge6$; Appendix~\ref{app:small} handles fewer processors.

In this section, we first solve a linear program that assigns each job a fractional duration and work value.
It relaxes the processor capacity constraints to a total-work bound and uses completion-time variables to bound every precedence path without listing the paths explicitly.

If two allocations give the same duration, the smaller allocation dominates the larger one.
For each job $j$, define the set of \emph{retained allocations}
\[
      \mathcal A_j=\{1\}\cup
      \{p\in\{2,\ldots,m\}:t_j(p)<t_j(p-1)\}.
\]
These are the smallest allocations attaining each duration, so the values
$t_j(p)$ for $p\in\mathcal A_j$ are distinct.
We preserve the original processor counts: consecutive elements of
$\mathcal A_j$ need not be consecutive integers.
The original processing-time table remains available for capping.
Restricting each job to its retained allocations preserves the optimum:
replacing a discarded allocation by the retained allocation of the same
duration only reduces its processor requirement.

\begin{definition}[Convex Duration-Work Frontier]
      \label{def:pub-frontier}
      The duration-work frontier $F_j$ is the continuous piecewise-linear
      interpolation through the points $(t_j(p),p\,t_j(p))$ for
      $p\in\mathcal A_j$.
      We call the table \emph{convex} if $F_j$ is convex on $[t_j(m),t_j(1)]$.
      A one-point frontier is convex by convention.
\end{definition}

\begin{figure}[tbp]
      \centering
      \begin{minipage}[c]{0.34\linewidth}
            \centering
            \begin{tabular}{ccc}
                  \toprule
                  $p$ & $t_j(p)$ & $p\,t_j(p)$ \\
                  \midrule
                  1   & 12       & 12          \\
                  2   & 8        & 16          \\
                  3   & 8        & 24          \\
                  4   & 6        & 24          \\
                  5   & $11/2$   & $55/2$      \\
                  \bottomrule
            \end{tabular}
      \end{minipage}\hfill
      \begin{minipage}[c]{0.63\linewidth}
            \centering
            \begin{tikzpicture}[x=0.65cm,y=0.19cm,font=\small]
                  \draw[->] (4.5,10) -- (13.1,10) node[below left] {duration};
                  \draw[->] (4.5,10) -- (4.5,31) node[above] {work};
                  \draw[thick] (5.5,27.5) -- (6,24) -- (8,16) -- (12,12);
                  \node[above right] at (9.8,14.2) {$F_j$};
                  \foreach \x/\y/\p in {5.5/27.5/5,6/24/4,8/16/2,12/12/1}{
                              \fill (\x,\y) circle (2pt);
                              \node[above right] at (\x,\y) {$p=\p$};
                        }
                  \draw[densely dashed,gray] (8,16) -- (8,24);
                  \draw[thick,fill=white] (8,24) circle (2.5pt);
                  \node[right,align=left] at (8.2,25.5)
                  {$p=3$\\discarded};
                  \draw[densely dashed,gray] (7,10) -- (7,20) -- (4.5,20);
                  \fill[gray] (7,20) circle (2pt);
                  \node[left] at (4.5,20) {$F_j(7)=20$};
                  \node[below] at (7,10) {$7$};
            \end{tikzpicture}
      \end{minipage}
      \caption{A processing-time table and its convex duration--work
            frontier. Allocation $p=3$ is dominated by $p=2$, which has the same
            duration and less work. The black points correspond to
            $\mathcal A_j=\{1,2,4,5\}$; the segments interpolate between them.
            For example, duration $7$ combines
            allocations $p=2$ and $p=4$ with equal weights and has fractional
            work $F_j(7)=20$, shown by the grey point.}
      \label{fig:duration-work-frontier}
\end{figure}
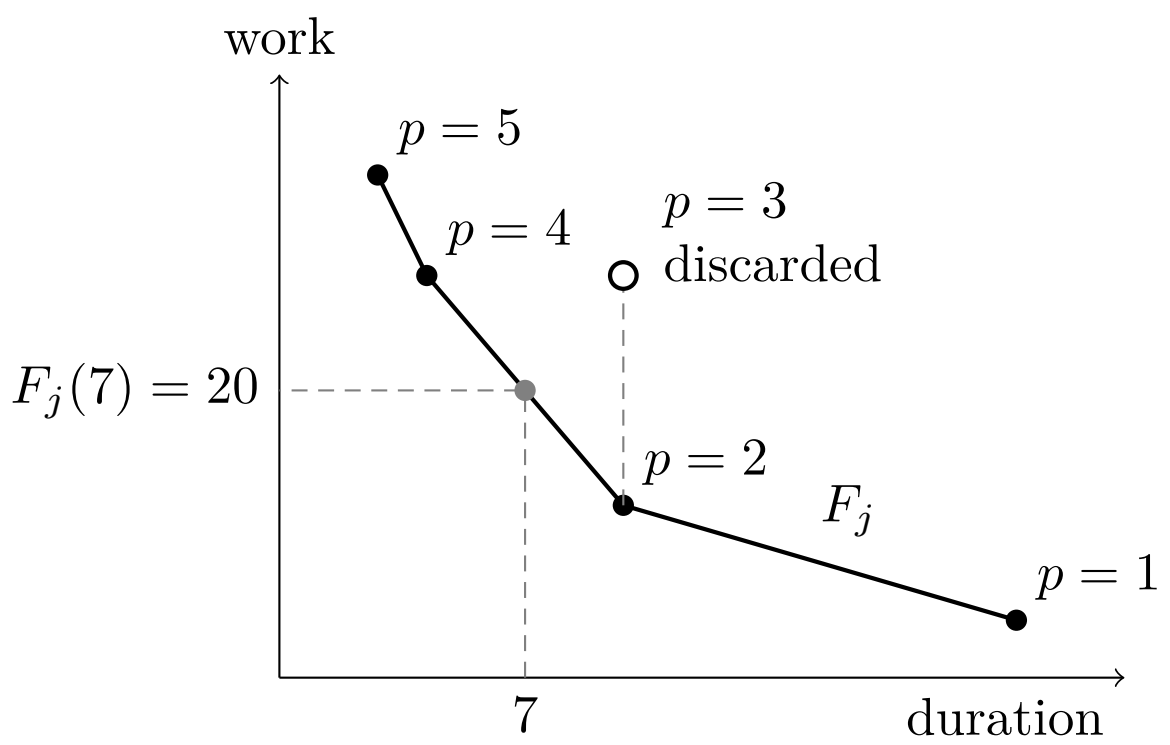

\Cref{fig:duration-work-frontier} illustrates the preprocessing and the resulting interpolation.
Points between retained durations represent fractional LP choices, rather than additional processor allocations.

We assume throughout that every frontier is convex.
Jansen and Zhang~\cite[Theorem 2.2]{jansen2012} derive this convexity from the stronger assumption of concave speedup.
Our model also allows interior duration plateaus, retaining the smallest allocation on each plateau.

For each job $j\in J$ and each retained allocation $p\in\mathcal A_j$, introduce a weight $\theta_{j,p}\ge0$.
Let $d_j$ and $w_j$ be the fractional duration and work, respectively, and let $c_j$ be a completion-time bound.
We use the assignment relaxation of Jansen and Zhang~\cite[Section~3.1, equation~(10)]{jansen2012}, written for the retained allocations:
\begin{align}
      \min\quad                           & T \nonumber                                                                                           \\
      \sum_{p\in\mathcal A_j}\theta_{j,p} & =1
                                          &                                                 & (j\in J),\nonumber                                  \\
      d_j                                 & =\sum_{p\in\mathcal A_j}\theta_{j,p}t_j(p)
                                          &                                                 & (j\in J),\nonumber                                  \\
      w_j                                 & =\sum_{p\in\mathcal A_j}\theta_{j,p}\,p\,t_j(p)
                                          &                                                 & (j\in J),\nonumber                                  \\
      c_j                                 & \ge d_j                                         &                    & (j\in J),\nonumber             \\
      c_j                                 & \ge c_i+d_j                                     &                    & ((i,j)\in E),\label{eq:pub-lp} \\
      c_j                                 & \le T                                           &                    & (j\in J),\nonumber             \\
      \sum_{j\in J}w_j                    & \le mT.\nonumber
\end{align}
The constraints $c_j\ge d_j$ are included explicitly, also for source jobs.
Together with the precedence constraints, they imply that the sum of the fractional durations along any directed path is at most $T$;
\Cref{pub-lp} proves this implication.
The LP has polynomial size.

Take an optimal LP solution and consider each job's duration $d_j$.
As a convex combination of retained durations, $d_j$ need not itself equal a retained duration.
If $d_j=t_j(s)$ for some $s\in\mathcal A_j$, replace the job's weights by weight one on allocation $s$ and call the job \emph{integral}.
Otherwise, there is a unique pair $s<f$ of consecutive elements of
$\mathcal A_j$ satisfying $t_j(s)>d_j>t_j(f)$.
Set
\[
      \theta_{j,s}=\frac{d_j-t_j(f)}{t_j(s)-t_j(f)},\qquad
      \theta_{j,f}=1-\theta_{j,s},
\]
and set all other weights to zero; call this job \emph{fractional}.
This replacement preserves $d_j$ and, by convexity, does not increase $w_j$, as proved in \Cref{pub-lp}.
Hence it remains optimal.
From now on, $w_j$ denotes the work after this replacement.

For a fractional job $j$, the allocations $s<f$ are its slower and faster
retained allocations, respectively. Write $\theta=\theta_{j,s}$, so that
$\theta_{j,f}=1-\theta$.
The corresponding durations and LP values satisfy
\begin{equation}
      \label{pub-main-adjacent}
      \begin{gathered}
            t_j(s)=\cdots=t_j(f-1)>t_j(f),\qquad 1\le s<f\le m,\\
            d_j=\theta t_j(s)+(1-\theta)t_j(f),\\
            w_j=\theta s\,t_j(s)+(1-\theta)f\,t_j(f),
            \qquad 0<\theta<1.
      \end{gathered}
\end{equation}
The slower duration plateau runs from allocation $s$ through $f-1$.
When considering several jobs, we write $s_j$ and $f_j$ for their retained allocations.
Monotonicity of work at allocations $f-1$ and $f$ gives
\begin{equation}
      \label{eq:pub-adjacency}
      \frac{t_j(f)}{t_j(s)}\ge\frac{f-1}{f}.
\end{equation}
This estimate is the reason for requiring consecutive retained allocations.

\subsection{Rounding, Capping, and List Scheduling}\label{sec:list-scheduling}

We construct a finite family of $N$ schedules from the same optimal LP solution, represented as above.
Choose thresholds $\alpha_i>1$ and cap parameters $2<r_i<3$, indexed by $i\in[N]$.
We leave these pairs and the family size $N$ unspecified until \Cref{sec:parameters}.
Define
\begin{equation}\label{eq:algorithm-caps}
      B_i=\left\lceil\frac{m}{r_i}\right\rceil,\qquad
      B_\star = \min_{i\in[N]}B_i,\qquad
      \alpha_i'=\max\left\{\alpha_i,1+\frac1{B_\star}\right\}.
\end{equation}

For a single run $i \in [N]$, write $\alpha=\alpha_i'$ and $B=B_i$.
We round and cap the LP solution as follows. For an integral job $j$, let $s$ be its retained allocation.
For a fractional job $j$, use its slower and faster retained allocations $s$ and $f$ from~\eqref{pub-main-adjacent}.
The final allocation is
\begin{align}\label{eq:job-allocation}
      p_j =
      \begin{cases}
            \min\{s,B\}, & \text{if $j$ is integral},                            \\
            \min\{s,B\}, & \text{if $j$ is fractional and } t_j(s)\le\alpha d_j, \\
            \min\{f,B\}, & \text{if $j$ is fractional and } t_j(s)>\alpha d_j.
      \end{cases}
\end{align}

Once the allocations are fixed, we use an event-driven list schedule.
At time zero and at each completion event, repeatedly start a ready job that fits on the idle processors until no such job remains.
A job is ready when all of its immediate predecessors have completed.
Fix any job ordering to break ties, making the rule deterministic.
If unfinished jobs remain and none is running, the precedence graph has a ready job, and that job fits because its allocation is at most $m$.
Thus the procedure always makes progress.
Every event starts a job or completes one, so there are polynomially many events.

For each run, we apply this rule to the rounded and capped jobs and return the shortest of the $N$ schedules.
\Cref{alg:threshold-rounding} summarises the complete algorithm.

\begin{algorithm}[H]
      \caption{Threshold Rounding with Processor Caps}
      \label{alg:threshold-rounding}
      \begin{algorithmic}[1]
            \Require An instance with $m\ge6$; pairs $(\alpha_i,r_i)_{i=1}^N$
            with $\alpha_i>1$ and $2<r_i<3$
            \State Remove dominated allocations and solve the LP~\eqref{eq:pub-lp}.
            \State Compute $B_i$, $B_\star$, and $\alpha_i'$ from~\eqref{eq:algorithm-caps}.
            \For{$i=1,\ldots,N$}
            \State Round jobs to allocations $p_j$ using~\eqref{eq:job-allocation}.
            \State Construct a list schedule $\mathcal S_i$ using durations $t_j(p_j)$.
            \EndFor
            \State \Return a schedule of minimum makespan among $\mathcal S_1,\ldots,\mathcal S_N$.
      \end{algorithmic}
\end{algorithm}

\section{Parameterised Analysis}\label{sec:analysis}

The analysis has three steps.
The LP bounds fractional path lengths and total work.
The list rule converts rounded durations and work into a makespan bound.
Finally, suitable weights combine these bounds over the threshold-cap family.
All LP quantities below refer to an optimal solution in which each job uses one retained allocation or a convex combination of two consecutive retained allocations.

The next lemma bounds the durations and work in an LP solution and justifies restricting each job to two consecutive retained allocations.

\begin{lemma}
      \label{pub-lp}
      The optimum of~\eqref{eq:pub-lp} satisfies $T\le\OPT$.
      Every feasible solution satisfies
      \begin{equation}\label{pub-main-lp-properties}
            \begin{gathered}
                  \sum_{j\in P}d_j\le T\quad\text{for every directed path }P,\\
                  t_j(1)\le w_j\le md_j,\qquad \sum_jw_j\le mT.
            \end{gathered}
      \end{equation}
      A solution can be adjusted, without changing its durations or increasing its work, so that each job uses one retained allocation or a convex combination of two consecutive retained allocations.
\end{lemma}

\begin{proof}
      A feasible schedule supplies an allocation per job, its completion times, and its makespan.
      Replacing an allocation by the retained allocation of the same duration only reduces processor usage, preserving feasibility.
      Integrating processor usage bounds total work by $m$ times the makespan, so these values are LP feasible and $T\le\OPT$.

      For a directed path $P=(j_1,\ldots,j_k)$, the LP constraints give $c_{j_1}\ge d_{j_1}$ and $c_{j_\ell}-c_{j_{\ell-1}}\ge d_{j_\ell}$ for $2\le\ell\le k$.
      Adding these inequalities yields
      \[
            \sum_{\ell=1}^k d_{j_\ell}
            \le c_{j_1}+\sum_{\ell=2}^k
            (c_{j_\ell}-c_{j_{\ell-1}})
            =c_{j_k}\le T.
      \]
      For each retained allocation $p\in\mathcal A_j$, monotonicity of work and $p\le m$ give $t_j(1)\le p\,t_j(p)\le m\,t_j(p)$.
      Taking the convex combination specified by the LP proves the individual work bounds; the total-work bound is an LP constraint.

      An LP duration $d_j$ is a convex combination of retained durations and need not itself be a retained duration.
      If $d_j=t_j(p)$ for some $p\in\mathcal A_j$, assign weight one to allocation $p$.
      Otherwise, choose the unique consecutive elements $s<f$ of
      $\mathcal A_j$ satisfying $t_j(s)>d_j>t_j(f)$ and assign weights
      \[
            \lambda=\frac{d_j-t_j(f)}{t_j(s)-t_j(f)},
            \qquad 1-\lambda
      \]
      to allocations $s$ and $f$, respectively.
      This preserves $d_j$.
      Let $F_j$ be the piecewise-linear duration-work function from \Cref{def:pub-frontier}.
      Its convexity gives, for the original weights,
      \[
            F_j(d_j)\le\sum_{p\in\mathcal A_j}\theta_{j,p}F_j(t_j(p))=w_j.
      \]
      The new weights attain $F_j(d_j)$, so work does not increase and all LP constraints remain valid.
      Finding the consecutive retained allocations and computing the new weights use polynomially many rational operations.
\end{proof}

We first bound the makespan in terms of a precedence path and the work of jobs outside that path.
We then derive a bound that no longer depends on the path.

Fix an integer cap $1\le B\le m$ and processor allocations $p_j\le B$, and use the list scheduling rule from \Cref{sec:list-scheduling}.
Write $C$ for the resulting makespan, and define
\[
      Q=m-B+1,\qquad U_j=p_jt_j(p_j),
\]
where $U_j$ is the work of job $j$ at its assigned allocation.

\begin{lemma}\label{pub-main-disjoint}
      There is a directed path $P$ such that
      \begin{equation}\label{pub-main-disjoint-equation}
            C\le\sum_{j\in P}t_j(p_j)+\frac{1}{Q}\sum_{j\notin P}U_j.
      \end{equation}
\end{lemma}
\begin{proof}
      Choose a last-finishing job and trace backwards, choosing for each non-source job an immediate predecessor whose completion time is latest.
      This gives a directed path $P$.
      Consider the intervals before the first path job starts and between successive executions of path jobs.
      Throughout each such interval, the next path job is ready: the first path job is a source, and the chosen predecessor of every later path job finishes no earlier than any of its other immediate predecessors.

      After the list-scheduling procedure has processed an event, any ready job that has not started does not fit.
      Since the next path job has allocation $p\le B$, at least $m-p+1\ge Q$ processors are busy.
      All jobs executing in the intervals under consideration lie outside $P$.
      By an area argument, the total length of these intervals is at most $\sum_{j\notin P}U_j/Q$.
      The remaining part of $[0,C]$ consists of the executions of the path jobs, proving the claim.
\end{proof}

The next lemma removes the path from the makespan bound.
We first define the two classes of jobs used throughout the remaining analysis. Recall that $B_\star=\min_i B_i$ is the smallest processor cap in \eqref{eq:algorithm-caps}.
Using the faster retained allocation $f_j$ of each fractional job, define
\begin{equation}\label{eq:thin-wide-partition}
      \mathcal T=\{j\in J: j\text{ is fractional and }f_j\le B_\star\},
      \qquad \mathcal W=J\setminus\mathcal T.
\end{equation}
We call the jobs in $\mathcal T$ \emph{thin} and those in $\mathcal W$ \emph{wide}. A thin job has both retained allocations at most $B_\star$.
The wide jobs are exactly the integral jobs and the fractional jobs with $f_j>B_\star$.
In particular, an integral job belongs to $\mathcal W$ regardless of its processor allocation.
The partition depends only on the LP solution and $B_\star$, so it is the same in every run.

Consider a single run with threshold $\alpha\ge1+1/B_\star$ and integer cap $B_\star\le B\le m$.
Write $\widehat p_j$ for the allocation selected by threshold rounding (before the min-operation in \eqref{eq:job-allocation}) and $p_j=\min\{\widehat p_j,B\}$ for the final allocation.
A job is \emph{capped} precisely when $\widehat p_j>B$.
Using the notation above for these final allocations, put
\[
      c=1-\frac1Q,\qquad
      W_{\mathcal W}=\sum_{j\in\mathcal W}w_j,\qquad
      U_{\mathcal T}=\sum_{j\in\mathcal T}U_j.
\]
Thus $W_{\mathcal W}$ is the LP work of the wide jobs, whereas $U_{\mathcal T}$ is the actual final work of the thin jobs in this run.

\begin{lemma}\label{pub-main-scalar}
      For the partition and rounding defined above, the resulting list schedule has makespan
      \begin{equation}\label{pub-main-scalar-equation}
            C\le c\alpha T+
            \max\left\{\frac1Q,\frac1B-\frac{c\alpha}{m}\right\}
            W_{\mathcal W}+\frac{U_{\mathcal T}}Q.
      \end{equation}
\end{lemma}
\begin{proof}
      A thin job satisfies $\widehat p_j\le B_\star\le B$, so it is never capped. For a fractional wide job, the notation in~\eqref{pub-main-adjacent} and the adjacency bound give
      \[
            \frac{t_j(s_j)}{d_j}\le\frac{t_j(s_j)}{t_j(f_j)}\le\frac{f_j}{f_j-1}
            \le1+\frac1{B_\star}\le\alpha.
      \]
      It therefore selects its slower allocation, whose work is at most $w_j$.
      An integral job has work $w_j$ before capping.
      Since capping cannot increase work, every wide job has final work $U_j\le w_j$.

      Define $v_j=U_j$ for thin jobs and $v_j=w_j$ for wide jobs.
      Then $U_j\le v_j$ in both cases.
      Apply \Cref{pub-main-disjoint}, replace each off-path work term by $v_j$, and write the off-path sum as the total sum minus the path sum:
      \begin{align*}
            C & \le\sum_{j\in P}t_j(p_j)+\frac1Q\sum_{j\notin P}v_j \\
              & =\sum_{j\in P}\left(t_j(p_j)-\frac{v_j}{Q}\right)
            +\frac1Q\sum_jv_j                                       \\
              & =\sum_{j\in P}\left(t_j(p_j)-\frac{v_j}{Q}\right)
            +\frac{W_{\mathcal W}+U_{\mathcal T}}Q.
      \end{align*}
      For a thin job $j\in P$, $v_j=p_jt_j(p_j)\ge t_j(p_j)$, and its allocation is unchanged by the cap.
      Threshold rounding therefore gives $t_j(p_j)\le\alpha d_j$, so its contribution in parentheses is at most $c\alpha d_j$.
      The same bound holds for a wide job whose selected allocation is at most $B$: its allocation is also unchanged, and $v_j=w_j\ge p_jt_j(p_j)\ge t_j(p_j)$.

      Every remaining path job is capped and wide.
      Its final allocation is $B$, so the final-work bound just proved gives $B\,t_j(B)\le w_j$.
      Its contribution in parentheses is therefore at most $(1/B-1/Q)w_j$.
      Since $w_j\le md_j$, this is at most
      \[
            c\alpha d_j+
            \max\left\{0,\frac1B-\frac1Q-\frac{c\alpha}{m}\right\}w_j.
      \]
      Sum over the path, apply its duration bound in \eqref{pub-main-lp-properties}, and bound the additional capped-path work by $W_{\mathcal W}$.
      Adding $W_{\mathcal W}/Q$ gives \eqref{pub-main-scalar-equation}.
\end{proof}

\subsection{Averaging across Runs}\label{sec:averaging}

We analyse the work obtained by rounding each job with the $N$ threshold choices.
Let $U_j(i)$ be the work of job $j$ after rounding with the original threshold $\alpha_i$, before capping.
Assign nonnegative weights $\nu_i$ to the threshold choices, with $\sum_i\nu_i=1$, and use the same weights for every job.
For a fixed job $j$, these weights define an average of its work across the $N$ threshold choices.
Our aim is to find conditions under which
\[
      \sum_i\nu_iU_j(i)\le\beta w_j\qquad\text{for every job }j,
\]
where $\beta>1$ is a target work factor.
Throughout this analysis, sums over $i$ range over $[N]$ unless explicitly restricted.

An integral job has work $w_j$ under every threshold.
For a fractional job $j$, use the notation in~\eqref{pub-main-adjacent}: threshold $\alpha_i$ selects the faster allocation precisely when $\alpha_i<t_j(s)/d_j$, and selects the slower allocation otherwise.
Its weighted work therefore depends only on the total weight $\sum_{i:\alpha_i<t_j(s)/d_j}\nu_i$ assigned to the faster choice.
We control this quantity by bounding the weight of thresholds below each value $y$ by a function $F(y)$.
The next two lemmas show that the function below gives a sufficient bound for every feasible pair of retained allocations and durations.
Raising the thresholds to $\alpha_i'$ and imposing the processor caps cannot increase work, so the resulting work guarantee also applies to the actual runs.

Write $D=\beta-1$ and, for $y>1$, define
\begin{equation}\label{pub-main-envelope-definition}
      \begin{split}
            A(y) & =\beta(1-1/y), \\
            F(y) & =\min\left\{
            \bigl(\sqrt{2A(y)}+\sqrt D\bigr)^2,
            \inf_{k\in\mathbb Z_{\ge1}}
            \left((k+2)A(y)+\frac Dk\right)\right\},
      \end{split}
\end{equation}
and set $F(1)=0$.
The function $F$ is nondecreasing, because $A(y)$ and every expression in the minimum are nondecreasing.

The expression in the next lemma arises from rearranging the target work inequality, as shown in the proof of Lemma~\ref{pub-main-work}.
Bounding it from below uniformly over the possible allocations and duration ratios yields the sufficient bound $F(y)$ on the weight assigned to faster choices.

\begin{lemma}[\faCut]\label{pub-main-envelope}
      For all $A,D>0$, every integer $b\ge2$, and every $1-1/b\le r<1$ with $br>1$, we have
      \begin{equation}\label{pub-main-envelope-inequality}
            \frac{A}{1-r}+\frac{D}{br-1}
            \ge\min\left\{(\sqrt{2A}+\sqrt D)^2,
            \inf_{k\in\mathbb Z_{\ge1}}\left((k+2)A+\frac Dk\right)
            \right\}.
      \end{equation}
\end{lemma}

We can now bound the weighted average of each job's work across the threshold choices.
The following lemma gives a sufficient condition for this average to be at most $\beta$ times the job's LP work, where $\beta>1$ is the target work factor used to define $F$.

\begin{lemma}\label{pub-main-work}
      Let $H>1$ satisfy $F(H)=1$. Suppose thresholds $\alpha_i>1$ and nonnegative weights $\nu_i$ satisfy
      \begin{equation}\label{pub-main-cdf-condition}
            \sum_i\nu_i=1,\qquad
            \sum_{i:\alpha_i<y}\nu_i\le F(y)\quad\text{for }1<y\le H.
      \end{equation}
      Then, for every job $j$,
      \begin{equation}\label{pub-main-work-target}
            \sum_i\nu_iU_j(i)\le\beta w_j.
      \end{equation}
      This inequality remains valid if any thresholds are increased.
\end{lemma}
\begin{proof}
      An integral job has work $w_j$ in every rounding, so its weighted
      work is $w_j\le\beta w_j$, as required by~\eqref{pub-main-work-target}.
      For a fractional job $j$, use~\eqref{pub-main-adjacent} and put
      \[
            r=\frac{t_j(f)}{t_j(s)},\qquad x=\frac{d_j}{t_j(s)},\qquad
            u=\sum_{i:\alpha_i<1/x}\nu_i.
      \]
      The slower allocation has work $s\,t_j(s)$, while the faster allocation has work $f\,t_j(f)=fr\,t_j(s)$.
      The faster allocation is chosen precisely for the thresholds counted by $u$; the remaining weight $1-u$ selects the slower allocation.
      Dividing the weighted rounded work by $t_j(s)$ therefore gives
      \begin{equation}\label{pub-main-normalized-rounded-work}
            \frac{\sum_i\nu_iU_j(i)}{t_j(s)}
            =(1-u)s+u\,fr=s+(fr-s)u.
      \end{equation}

      In the LP representation~\eqref{pub-main-adjacent}, the weights on the slower and faster allocations are $\theta$ and $1-\theta$.
      Dividing the LP duration equation by $t_j(s)$ yields
      \[
            x=\theta+(1-\theta)r=1-(1-\theta)(1-r).
      \]
      Since $r<1$, the LP weight on the faster allocation is $1-\theta=(1-x)/(1-r)$.
      Using these weights to compute the LP work and then dividing by $t_j(s)$, we obtain
      \begin{equation}\label{pub-main-normalized-lp-work}
            \begin{aligned}
                  w_j & =\theta s\,t_j(s)+(1-\theta)f\,t_j(f), \\
                  \Leftrightarrow \frac{w_j}{t_j(s)}
                      & =\theta s+(1-\theta)fr
                  =s+(fr-s)\frac{1-x}{1-r}.
            \end{aligned}
      \end{equation}
      Equations~\eqref{pub-main-normalized-rounded-work} and~\eqref{pub-main-normalized-lp-work} express both normalised work quantities as weighted averages of the same two values, $s$ and $fr$.
      The faster allocation has weight $u$ after rounding and $(1-x)/(1-r)$ in the LP; in each case, the slower allocation receives the remaining weight.

      Since $s<f$, monotonicity of work gives $s\,t_j(s)\le f\,t_j(f)$, and hence $fr\ge s$.
      If $fr=s$, both retained allocations have the same work.
      Therefore
      \[
            \sum_i\nu_iU_j(i)=w_j=s\,t_j(s)\le\beta w_j.
      \]
      Thus the target inequality~\eqref{pub-main-work-target} holds regardless of the rounding choices.

      Now suppose $fr>s\ge1$.
      Our goal is $\sum_i\nu_iU_j(i)\le\beta w_j$, as stated in~\eqref{pub-main-work-target}.
      Dividing this inequality by $t_j(s)>0$ and substituting \eqref{pub-main-normalized-rounded-work} and~\eqref{pub-main-normalized-lp-work} shows that it is equivalent to
      \begin{equation}\label{pub-main-normalized-work-target}
            s+(fr-s)u\le\beta\left(s+(fr-s)\frac{1-x}{1-r}\right).
      \end{equation}
      Subtracting $s$ from both sides of~\eqref{pub-main-normalized-work-target} and dividing by $fr-s>0$ gives, with $D=\beta-1$,
      \begin{equation}\label{pub-main-work-sufficient}
            u\le\frac{\beta(1-x)}{1-r}+D\frac{s}{fr-s}.
      \end{equation}
      Since $s\ge1$, we have $s/(fr-s)\ge1/(fr-1)$.
      Also $r\ge1-1/f$ by~\eqref{eq:pub-adjacency}.
      If $1/x\le H$, condition~\eqref{pub-main-cdf-condition} and \Cref{pub-main-envelope}, with $b=f$ and $A=\beta(1-x)=A(1/x)$, prove~\eqref{pub-main-work-sufficient}.
      If $1/x>H$, apply the same lemma with $b=f$ and $A=A(H)$.
      Its right-hand side equals $F(H)=1$, and $\beta(1-x)>A(H)$.
      Consequently the right-hand side of \eqref{pub-main-work-sufficient} is at least $1\ge u$.
      Finally, increasing a threshold cannot increase the rounded work.
\end{proof}

We now combine the work bound with a weighted average of the constructed schedules' makespans.
The weights $\nu_i$ average job work, while $\pi_i$ average schedule makespans.
The algorithm returns the shortest schedule, whose makespan is at most this average.

Fix rational thresholds $\alpha_i>1$ and cap parameters $2<r_i<3$, indexed by $i\in[N]$.
Let $\pi_i,\nu_i\ge0$ be rational weights with $\sum_i\pi_i=\sum_i\nu_i=1$.
For a real number $z$, write $(z)_+=\max\{z,0\}$, and define
\[
      \begin{gathered}
            k_i=\frac{r_i}{r_i-1},\qquad r_{\max}=\max_i r_i,\qquad
            \mu=\sum_i\pi_i\alpha_i,\qquad
            \delta_3=\sum_i\pi_i(4/3-\alpha_i)_+,\\
            V=\sum_i\pi_i\max\{\alpha_i+k_i,r_i\}.
      \end{gathered}
\]

\begin{lemma}\label{pub-main-lift}
      Assume that $\sum_i\nu_iU_j(i)\le\beta w_j$ for every job $j$,
      and that a number $\kappa>0$ satisfies
      \begin{equation}\label{pub-main-lift-conditions}
            \begin{gathered}
                  \pi_i k_i\le\kappa\nu_i\quad\text{for every }i,\qquad
                  3(r_{\max}-1)\delta_3\le\mu,\\
                  \frac{4(r_{\max}-1)}{3r_{\max}-2}\le\min_i(r_i-k_i).
            \end{gathered}
      \end{equation}
      For every $m\ge6$, the algorithm of \Cref{pub-main-section} with these threshold-cap pairs returns a schedule of makespan at most
      \[
            \max\{\mu+\beta\kappa,V\}\,T.
      \]
\end{lemma}

\begin{proof}
      For each run $i$, we apply the makespan bound \eqref{pub-main-scalar-equation} from \Cref{pub-main-scalar}.
      We then average these bounds with weights $\pi_i$.
      Let $B_i$, $B_\star$, and $\alpha_i'$ be as in \eqref{eq:algorithm-caps}, and abbreviate
      \[
            Q_i=m-B_i+1,\qquad c_i=1-\frac1{Q_i}.
      \]
      Since $m\ge6$ and $r_{\max}<3$, we have $B_\star\ge3$.

      Use the fixed partition $\mathcal T,\mathcal W$ from \eqref{eq:thin-wide-partition}.
      Since $B_i\ge B_\star$ and $\alpha_i'\ge1+1/B_\star$, \Cref{pub-main-scalar} applies to every run with this same partition.
      In particular, thin jobs are uncapped and wide jobs have final work at most $w_j$ in every run.

      Since $B_i=\lceil m/r_i\rceil$, we have
      $m/r_i\le B_i<m/r_i+1$. Thus
      \begin{equation}\label{pub-main-cap-bounds}
            \frac m{B_i}\le r_i,\qquad
            Q_i=m-B_i+1>m\left(1-\frac1{r_i}\right)=\frac m{k_i},
      \end{equation}
      and hence $m/Q_i < k_i$.
      We also have $B_\star=\lceil m/r_{\max}\rceil$, so $m-B_\star\le(r_{\max}-1)B_\star$.
      Since $B_i\ge B_\star$, we have $Q_i\le m-B_\star+1$ and therefore
      \begin{equation}\label{pub-main-cap-coefficient}
            c_i\le c_\star:=\frac{m-B_\star}{m-B_\star+1}.
      \end{equation}

      To prepare for averaging the makespan bounds, set
      \[
            x=\frac{\sum_{j\in\mathcal W}w_j}{mT},\qquad
            y=\frac{\sum_{j\in\mathcal T}w_j}{mT}.
      \]
      Then $x,y\ge0$ and $x+y\le1$. In the makespan bound
      \eqref{pub-main-scalar-equation}, the duration term and the
      wide-job work term, after division by $T$, can be rewritten as
      \begin{equation}\label{pub-main-makespan-rewriting}
            \begin{aligned}
                   & c_i\alpha_i'
                  +x\max\{m/Q_i,m/B_i-c_i\alpha_i'\} \\
                   & \qquad=(1-x)c_i\alpha_i'
                  +x\max\{c_i\alpha_i'+m/Q_i,m/B_i\}.
            \end{aligned}
      \end{equation}
      The following two claims will bound the weighted averages of the
      coefficients of $1-x$ and $x$ by $\mu$ and $V$, respectively.

      \begin{claim}\label{pub-main-threshold-claim}
            The weighted average of the threshold coefficients satisfies
            \[
                  \sum_i\pi_i c_i\alpha_i'\le\mu.
            \]
      \end{claim}
      \begin{proof}[Proof of \Cref{pub-main-threshold-claim}]
            \renewcommand{\qedsymbol}{\ensuremath{\blacksquare}}
            Raising the thresholds from $\alpha_i$ to
            $\alpha_i'=\max\{\alpha_i,1+1/B_\star\}$ increases their weighted
            average by
            \[
                  \Delta=\sum_i\pi_i(1+1/B_\star-\alpha_i)_+,
                  \qquad \sum_i\pi_i\alpha_i'=\mu+\Delta.
            \]
            We show that the factors $c_i<1$ compensate for this increase.
            Because $B_\star\ge3$ and $\alpha_i>1$,
            \[
                  \begin{aligned}
                        B_\star\Delta
                         & =\sum_i\pi_i\bigl(1-B_\star(\alpha_i-1)\bigr)_+ \\
                         & \le\sum_i\pi_i\bigl(1-3(\alpha_i-1)\bigr)_+
                        = 3 \sum_i\pi_i(4/3-\alpha_i)_+
                        =3\delta_3.
                  \end{aligned}
            \]
            Combining $m-B_\star\le(r_{\max}-1)B_\star$ with $B_\star\Delta\le3\delta_3$ and the second condition in~\eqref{pub-main-lift-conditions}, we obtain
            \[
                  (m-B_\star)\Delta
                  \le(r_{\max}-1)B_\star\Delta
                  \le3(r_{\max}-1)\delta_3\le\mu.
            \]
            Using~\eqref{pub-main-cap-coefficient}, we therefore obtain
            \begin{equation}\label{pub-main-clamp-average}
                  \begin{aligned}
                        \sum_i\pi_i c_i\alpha_i'
                         & \le c_\star(\mu+\Delta)                        \\
                         & =\frac{(m-B_\star)(\mu+\Delta)}{m-B_\star+1}   \\
                         & =\frac{(m-B_\star+1)\mu+(m-B_\star)\Delta-\mu}
                            {m-B_\star+1}  \\
                         & =\mu+\frac{(m-B_\star)\Delta-\mu}{m-B_\star+1}
                        \le\mu.
                  \end{aligned}
            \end{equation}
            This proves the claim, since $(m-B_\star)\Delta\le\mu$.
      \end{proof}

      \begin{claim}\label{pub-main-wide-claim}
            For every run $i\in[N]$,
            \begin{equation}\label{pub-main-wide-pointwise}
                  \max\{c_i\alpha_i'+m/Q_i,m/B_i\}
                  \le\max\{\alpha_i+k_i,r_i\}.
            \end{equation}
      \end{claim}
      \begin{proof}[Proof of \Cref{pub-main-wide-claim}]
            \renewcommand{\qedsymbol}{\ensuremath{\blacksquare}}
            If $\alpha_i\ge1+1/B_\star$, the threshold is unchanged, so
            $\alpha_i'=\alpha_i$. Since $c_i\le1$ and $m/Q_i\le k_i$,
            \[
                  c_i\alpha_i'+m/Q_i
                  =c_i\alpha_i+m/Q_i\le\alpha_i+k_i,
                  \qquad m/B_i\le r_i.
            \]
            Taking the maximum proves~\eqref{pub-main-wide-pointwise} in this case.
            If $\alpha_i<1+1/B_\star$, the threshold is raised to
            $\alpha_i'=1+1/B_\star$.
            In this case we show that both entries of the maximum are at most
            $r_i$. Since $m/Q_i\le k_i$, it suffices to show that
            $c_i\alpha_i'\le r_i-k_i$. By~\eqref{pub-main-cap-coefficient},
            \[
                  \begin{aligned}
                        c_i\alpha_i'
                         & \le\frac{m-B_\star}{m-B_\star+1}
                        \left(1+\frac1{B_\star}\right)          \\
                         & \le\frac{(r_{\max}-1)(B_\star+1)}
                              {(r_{\max}-1)B_\star+1}     \\
                         & \le\frac{4(r_{\max}-1)}{3r_{\max}-2}
                        \le r_i-k_i.
                  \end{aligned}
            \]
            The second inequality uses $m-B_\star\le(r_{\max}-1)B_\star$.
            For the third, the preceding fraction equals
            $1+(r_{\max}-2)/((r_{\max}-1)B_\star+1)$, which decreases with
            $B_\star$ because $r_{\max}>2$; we use $B_\star\ge3$.
            The last inequality is the final condition
            in~\eqref{pub-main-lift-conditions}.
            Thus $c_i\alpha_i'+m/Q_i\le r_i$ whenever the threshold is raised.
            Since also $m/B_i\le r_i$, both entries of the maximum are at most
            $r_i$, proving~\eqref{pub-main-wide-pointwise} in the remaining case.
      \end{proof}

      We now bound the average contribution of thin jobs and then average
      the makespan bounds.
      Let $U_{\mathcal T,i}$ be the final work of the thin jobs in run
      $i$. Recall that $U_j(i)$ is the work of job $j$ after rounding with
      the original threshold $\alpha_i$, before capping. Raising the
      threshold to $\alpha_i'$ can only leave the selected allocation
      unchanged or switch it from the faster to the slower allocation.
      Since work is nondecreasing in the allocation, this cannot increase
      a job's work. Thin jobs are uncapped, so
      $U_{\mathcal T,i}\le\sum_{j\in\mathcal T}U_j(i)$.
      Multiplying by $\nu_i\ge0$, summing over runs, and then applying
      the assumed work bound to each job in the fixed set $\mathcal T$ gives
      \[
            \sum_i\nu_iU_{\mathcal T,i}
            \le\sum_{j\in\mathcal T}\sum_i\nu_iU_j(i)
            \le\beta\sum_{j\in\mathcal T}w_j.
      \]
      The makespan bounds use the schedule weights $\pi_i$ and the
      coefficients $1/Q_i$. We relate them to this work average as follows:
      \begin{equation}\label{pub-main-weighted-thin-work}
            \begin{aligned}
                  \sum_i\pi_i\frac{U_{\mathcal T,i}}{Q_i}
                   & \le\frac1m\sum_i\pi_i k_iU_{\mathcal T,i}         \\
                   & \le\frac{\kappa}{m}\sum_i\nu_iU_{\mathcal T,i}    \\
                   & \le\frac{\beta\kappa}{m}\sum_{j\in\mathcal T}w_j.
            \end{aligned}
      \end{equation}
      The first inequality uses $1/Q_i\le k_i/m$, which follows from
      the cap bound~\eqref{pub-main-cap-bounds}. The second uses
      $\pi_i k_i\le\kappa\nu_i$, the first condition
      in~\eqref{pub-main-lift-conditions}; it converts the schedule weights
      into the work weights. Both inequalities apply term by term because
      the weights and work values are nonnegative. The last inequality
      applies the bound on $\sum_i\nu_iU_{\mathcal T,i}$ just proved.

      Let $C_i$ be the makespan of the schedule from run $i$.
      Dividing~\eqref{pub-main-scalar-equation} by $T$ and using the
      rewriting~\eqref{pub-main-makespan-rewriting} gives, for every run,
      \[
            \frac{C_i}{T}
            \le(1-x)c_i\alpha_i'
            +x\max\{c_i\alpha_i'+m/Q_i,m/B_i\}
            +\frac{U_{\mathcal T,i}}{Q_iT}.
      \]
      The algorithm returns the shortest of these schedules. Since
      $\pi_i\ge0$ and $\sum_i\pi_i=1$, its makespan is at most their
      weighted average: $\min_i C_i\le\sum_i\pi_i C_i$.
      Moreover, $x$ is the same in every run because it is defined using
      the LP work of the fixed set $\mathcal W$. Multiplying the
      per-run bounds by $\pi_i$ and summing therefore gives
      \begin{align*}
            \frac{\min_iC_i}{T}
             & \le\sum_i\pi_i\frac{C_i}{T}      \\
             & \le(1-x)\sum_i\pi_i c_i\alpha_i'
            +x\sum_i\pi_i\max\{c_i\alpha_i'+m/Q_i,m/B_i\}+\frac1T\sum_i\pi_i\frac{U_{\mathcal T,i}}{Q_i}.
      \end{align*}
      By~\eqref{pub-main-weighted-thin-work} and the definition of $y$,
      the last term satisfies
      \[
            \frac1T\sum_i\pi_i\frac{U_{\mathcal T,i}}{Q_i}
            \le\frac{\beta\kappa}{mT}\sum_{j\in\mathcal T}w_j
            =\beta\kappa y.
      \]
      \Needspace{8\baselineskip}
      Claim~\ref{pub-main-threshold-claim} bounds the first weighted
      coefficient by $\mu$. Claim~\ref{pub-main-wide-claim}, together
      with $V=\sum_i\pi_i\max\{\alpha_i+k_i,r_i\}$, bounds the second
      by $V$. Since $x\ge0$ and $1-x\ge0$, we can apply these bounds
      to obtain
      \begin{align*}
            \frac{\min_iC_i}{T}
             & \le\mu(1-x)+Vx+\beta\kappa y  \\
             & \le(\mu+\beta\kappa)(1-x)+Vx  \\
             & \le\max\{\mu+\beta\kappa,V\}.
      \end{align*}
      The second inequality uses $y\le1-x$ and $\beta\kappa>0$.
      The last follows because
      $(1-x)(\mu+\beta\kappa)+xV$ is a convex combination of
      $\mu+\beta\kappa$ and $V$, as $0\le x\le1$.
      The averaging is valid even if the paths in
      Lemma~\ref{pub-main-disjoint} differ between runs, since
      \eqref{pub-main-scalar-equation} eliminates each path before the
      schedules are averaged.
\end{proof}

\section{Choosing and Certifying the Parameters}
\label{sec:parameters}

We choose the threshold--cap family once, independently of the scheduling instance, to minimise the guarantee supplied by \Cref{pub-main-lift}.

The paper gives a self-contained proof of the guarantee for the chosen parameters, including the rational construction, mathematical arguments, exact sums, and rational bounds in Appendix~\ref{app:certificate}.
We used Python scripts to find the parameters and perform the finite arithmetic checks specified there; they are provided at \url{https://github.com/Felioh/moldable-precedence-scheduling} to make those calculations reproducible.

We minimise the objective
\begin{equation}\label{pub-main-parameter-objective}
      \max\{\mu+\beta\kappa,V\}.
\end{equation}
The choices are $\beta>1$, any finite number $N$ of pairs $(\alpha_i,r_i)$ with $\alpha_i>1$ and $2<r_i<3$, probability weights $\nu_i,\pi_i$, and $\kappa>0$.
They must satisfy the cumulative-weight condition~\eqref{pub-main-cdf-condition} and the domination and finite-processor conditions~\eqref{pub-main-lift-conditions}.
The quantities $\mu$ and $V$ are defined before \Cref{pub-main-lift}: $\mu+\beta\kappa$ bounds the thin-job contribution and $V$ bounds the wide-job contribution.
Thus every feasible rational choice certifies an approximation guarantee.

To find a good choice, we search a family described by three scalars $(t,\xi,\eta)$.
For $0<t\le1/5$, we set $\beta=1+t^2$, place thresholds uniformly between $1$ and $H=2(1+t^2)/(1+t)^2$, and use increments of the envelope $F$ as work weights.
The cap curve~\eqref{pub-main-cap-function} trades path inflation against capacity, while schedule weights proportional to $\nu_i(1-1/r_i)$ enforce domination.
The Python search varies $t$ and $\eta$ and adjusts $\xi$ to balance the two objective terms.
This is a numerical search within a restricted family, not a proof of global optimality.

We then replace the search parameters by rational numbers, round the cumulative work weights down and the caps up, and adjust the schedule weights to preserve domination.
Appendix~\ref{app:certificate} gives the full construction, using $N=65536$, and its exact sums.
The program \path{verify_main_certificate.py} checks every sufficient condition using integer and rational arithmetic; the resulting guarantee does not depend on numerical convergence or floating-point tolerances.
\begin{lemma}\label{pub-main-certificate}
      The rational construction in Appendix~\ref{app:certificate} satisfies all the conditions of \Cref{pub-main-lift} and has
      \[
            V\approx2.802670863310,\qquad
            \mu+\beta\kappa\approx2.802670869270.
      \]
      Both coefficients lie strictly between $11/4$ and $\rho=2802671/10^6=2.802671$.
\end{lemma}

\section{The Approximation Guarantee and Running Time}
\label{sec:final-guarantee}

We now apply the parameterised bound to the rational family in \Cref{sec:parameters}.
The search is no longer part of the algorithm: the resulting constants are fixed independently of the input.

\begin{proof}[Proof of \Cref{pub-main-theorem}]
      Let $m\ge6$.
      By \Cref{pub-main-grid-validity}, the weights $\nu_i$ give the average-work bound with factor $\beta$.
      Equation~\eqref{pub-main-domination} supplies the domination condition for the schedule weights.
      \Cref{pub-main-certificate} verifies the remaining hypotheses of \Cref{pub-main-lift}.
      Moreover, $\alpha_i+k_i\le r_i$ for every $i$, so $V=\sum_i\pi_i r_i$.
      Thus the returned schedule satisfies
      \[
            \Cmax\le\max\left\{\mu+\beta\kappa,\sum_i\pi_i r_i\right\}T
            \le\rho T\le\rho\OPT.
      \]
      For $m\le5$, \Cref{pub-main-small} in Appendix~\ref{app:small} supplies a schedule of length at most $(11/4)T<\rho\OPT$.
      This proves the guarantee for every positive number of processors.

      It remains to justify polynomial running time.
      The LP has polynomial size in the binary encoding of the explicit processing-time tables and the precedence graph.
      A rational optimal solution of polynomial bit length can be computed in polynomial time; replacing each job's support by consecutive retained allocations uses rational interpolation. The threshold-cap family has the constant size $N=65536$, and all its numbers have constant bit length.
      In each run, rounding and capping require polynomially many comparisons.
      The list schedule has at most one start and one completion event per job, and scanning the ready jobs at each event suffices to implement it.
      Its processing times and event times are rational numbers of polynomial bit length.
      The small-$m$ procedures also use only rounding and list scheduling, or a topological ordering.
      Hence the complete algorithm is deterministic and polynomial.
\end{proof}

\section{Declaration of Generative AI Use}
The authors developed the approach to improving the framework used in Jansen and Zhang~\cite{jansen2012} before using generative AI.
GPT-6 Astra provided substantial assistance with the analysis: starting from the authors' ideas, it derived an initial $3.004$-approximation guarantee on September 25, 2026.
Through successive iterations, the algorithm and its analysis were refined to the final version presented in this paper.
The authors extensively revised the presentation and take full responsibility for the paper's content and correctness.

\printbibliography

\appendix
\section{At Most Five Processors}\label{app:small}

This appendix supplies the small-processor cases omitted from the main
algorithm. The same LP lower bound applies. For $m\le2$, execute all jobs
sequentially on one processor in a topological order. For
$3\le m\le5$, round every fractional job to its slower endpoint, keep
every integral allocation, cap all allocations at two, and apply the list rule.

\begin{lemma}\label{pub-main-small}
      For $m\le5$, the procedures above return a schedule of length at most
      $(11/4)T$.
\end{lemma}
\begin{proof}
      If $m\le2$, the
      sequential schedule has length
      \[
            \sum_jt_j(1)\le\sum_jw_j\le mT\le2T.
      \]
      Now let $3\le m\le5$. Rounding every fractional job to its slower
      endpoint gives duration at most $2d_j$, because
      $t_j(s)/d_j\le f/(f-1)\le2$, and work at most $w_j$. Capping at two
      preserves the work bound. Put $Q=m-1$ and apply
      Lemma~\ref{pub-main-disjoint}, bounding every final work by $w_j$:
      \[
            C\le\sum_{j\in P}\left(t_j(p_j)-\frac{w_j}{Q}\right)
            +\frac1Q\sum_jw_j.
      \]
      A job assigned one processor was not capped, and its path contribution is
      at most $2(1-1/Q)d_j$. A job assigned two processors satisfies
      $2t_j(p_j)\le w_j\le md_j$, and its path contribution is at most
      $m(1/2-1/Q)d_j$, since $Q\ge2$. Using the path and work bounds in
      \eqref{pub-main-lp-properties}, we obtain
      \[
            \frac CT\le
            \max\left\{2\left(1-\frac1Q\right),
            m\left(\frac12-\frac1Q\right)\right\}
            +\frac mQ
            =\max\left\{2+\frac{m-2}{m-1},\frac m2\right\}
            \le\frac{11}{4}.
      \]
\end{proof}

\section{Proof of \Cref{pub-main-envelope}}\label{app:envelope}

\begin{proof}[Proof of \Cref{pub-main-envelope}]
      We minimise the left-hand side over its admissible interval and compare
      the minimum with one of the terms defining the envelope. Set $u=1-r$.
      Then $0<u\le1/b$, and the expression becomes
      \[
            \frac A{u}+\frac D{b-1-bu}.
      \]
      For positive denominators, the Cauchy--Schwarz inequality gives
      \[
            \frac A{u}+\frac D{b-1-bu}
            \ge\frac{(\sqrt{bA}+\sqrt D)^2}{b-1}.
      \]
      For $b=2$, the right-hand side is
      $(\sqrt{2A}+\sqrt D)^2$, which is one of the required envelope terms.

      Now let $b\ge3$. The expression is strictly convex in $u$ and tends
      to infinity as $u$ decreases to zero. If its minimum on $(0,1/b]$ is
      at $u=1/b$, that minimum is
      \[
            bA+\frac D{b-2},
      \]
      the integer envelope term with $k=b-2$.

      Otherwise the minimum is attained in the interior. Solving the
      derivative equation shows that its value and the condition for an
      interior minimiser are
      \[
            \frac{(\sqrt{bA}+\sqrt D)^2}{b-1},
            \qquad A<\frac{D}{b(b-2)^2}.
      \]
      In this case we compare with the integer term for $k=b-1$. The
      difference is
      \[
            \frac{(\sqrt{bA}+\sqrt D)^2}{b-1}
            -\left((b+1)A+\frac D{b-1}\right)
            =\frac{2\sqrt{bAD}-(b^2-b-1)A}{b-1}.
      \]
      The interior condition gives
      $2\sqrt{bAD}\ge2b(b-2)A$, and
      $2b(b-2)-(b^2-b-1)=b^2-3b+1>0$ for $b\ge3$.
      The difference is therefore nonnegative. In every case one of the
      terms in the stated envelope is at most the minimum of the left-hand
      side, proving the lemma.
\end{proof}

\section{Exact Verification of the Parameter Choice}
\label{app:certificate}

This appendix describes and justifies the calculations performed by the provided Python script \path{verify_main_certificate.py}, which verifies the fixed parameter choice using exact integer and rational arithmetic.
It supplies the algebraic details omitted from \Cref{sec:parameters}.

\subsection{The Rational Construction}

The numerical search uses the cap curve
\begin{equation}\label{pub-main-cap-function}
      r(\alpha)=\max\left\{\frac{30}{11},
      \frac{\alpha+2+\sqrt{\alpha^2+4}}2,
      \sqrt{\xi+\frac{\xi-\alpha}{\eta}}\right\}.
\end{equation}
The middle term ensures $\alpha+r/(r-1)\le r$. The last term
is chosen to balance the objective coefficients $\mu+\beta\kappa$ and $V$;
the first supplies slack for the finite-processor conditions.
For the exact construction, fix
\begin{equation}\label{pub-main-constants}
      \begin{gathered}
            t=\frac{1137}{6389},\qquad \xi=\frac{112084913}{40000000},\qquad
            \eta=\frac{484}{1327},\\
            \beta=1+t^2=\frac{42112090}{40819321},\qquad H=\frac{2(1+t^2)}{(1+t)^2}=\frac{21056045}{14160169},\\
            N=65536,\qquad M=10^{10},\qquad S=10^8.
      \end{gathered}
\end{equation}
The constants $N$, $M$, and $S$ control the threshold grid and the two
rational rounding precisions. They are independent of the input instance.
For $1\le i\le N$, set
\[
      \alpha_i=1+\frac{i}{N}(H-1),\qquad A_i=\beta(1-1/\alpha_i).
\]
To keep the cumulative work weights below $F$, round them down. Set
$z_0=0$, $z_N=M$, and, for $1\le i<N$,
\begin{equation}\label{pub-main-cdf-grid}
      \begin{split}
            z_i=\min\Biggl\{ & M,\quad
            \min_{1\le k\le64}\left\lfloor
            M\left((k+2)A_i+\frac Dk\right)\right\rfloor, \\
                             & \lfloor M(2A_i+D)\rfloor+
            \left\lfloor\sqrt{8M^2A_iD}\right\rfloor\Biggr\}.
      \end{split}
\end{equation}
Here $D=\beta-1$. Let $\nu_i=(z_i-z_{i-1})/M$. The cutoff $64$ is
valid for these particular constants, as justified below.

\begin{lemma}\label{pub-main-grid-validity}
      The weights $\nu_i$ are nonnegative, sum to one, and satisfy
      \eqref{pub-main-cdf-condition}. Consequently, they give average rounded
      work at most $\beta w_j$ for each job.
\end{lemma}

The proof below checks that the rounded cumulative weights are
nondecreasing and remain below the envelope.

Let $r_i$ be the smallest multiple of $1/S$ that is at least
$r(\alpha_i)$ from~\eqref{pub-main-cap-function}. To preserve the domination condition
after this rounding, now use the rational values
\begin{equation}\label{pub-main-weights}
      k_i=\frac{r_i}{r_i-1},\qquad
      h_i=\frac{\lfloor S(1-1/r_i)\rfloor}{S},\qquad
      Z=\sum_i\nu_i h_i,\qquad
      \pi_i=\frac{\nu_i h_i}{Z}.
\end{equation}
The downward rounding of $h_i$ ensures $0<h_i\le1/k_i$. Therefore
$Z>0$, the weights $\pi_i$ sum to one, and
\begin{equation}\label{pub-main-domination}
      \pi_i k_i\le\frac{\nu_i}{Z}.
\end{equation}
Thus we may still use $\kappa=1/Z$. Moreover,
$\alpha_i+k_i\le r_i$, so $V=\sum_i\pi_i r_i$ also remains valid.

\subsection{The Envelope Endpoint and the Cumulative Weights}

First consider the general search parameter $0<t\le1/5$. With
$\beta=1+t^2$ and $H=2\beta/(1+t)^2$, we have
\[
      2A(H)=(1-t)^2,\qquad D=t^2.
\]
Thus the square-root entry of $F(H)$ equals $(1-t+t)^2=1$.
The difference between consecutive integer entries is
\[
      ((k+3)A+D/(k+1))-((k+2)A+D/k)
      =A-\frac{D}{k(k+1)}.
\]
At $A=A(H)$ this is positive for every $k\ge1$, because
$A(H)\ge8/25$ and $D\le1/25$. The smallest integer entry is therefore
\[
      3A(H)+D=1+\frac{(1-5t)(1-t)}2\ge1.
\]
This proves $F(H)=1$ throughout the search range. For the constants
used in the paper, $2A(H)=(5252/6389)^2$ and $D=(1137/6389)^2$.

\begin{proof}[Proof of Lemma~\ref{pub-main-grid-validity}]
      Every expression defining $z_i$ is nondecreasing in $A_i$, and $A_i$
      is nondecreasing in $i$. Taking minima and floors preserves this
      property. Since every $z_i$ is at most $M$, the final choice $z_N=M$
      preserves it as well. Consequently $\nu_i\ge0$, and telescoping gives
      $\sum_i\nu_i=(z_N-z_0)/M=1$.

      We next justify restricting the integer minimum to $1\le k\le64$.
      Direct substitution of the fixed constants gives
      $A_1>D/(64\cdot65)$. For every $i\ge1$ and every $k\ge64$, the
      difference between consecutive integer entries is therefore positive:
      \[
            A_i-\frac{D}{k(k+1)}>0.
      \]
      Hence no omitted entry is smaller than the entry with $k=64$.
      The other rounded entry satisfies
      \[
            \lfloor M(2A_i+D)\rfloor+
            \left\lfloor\sqrt{8M^2A_iD}\right\rfloor
            \le M\bigl(\sqrt{2A_i}+\sqrt D\bigr)^2.
      \]
      It follows that $z_i/M\le F(\alpha_i)$ for $i<N$; at $i=N$ the
      same statement holds with equality because $F(H)=1$.

      Finally fix $1<y\le H$. If no threshold is strictly below $y$, the
      required cumulative bound is immediate. Otherwise let $\ell$ be the
      largest index with $\alpha_\ell<y$. Then $\ell<N$, and
      \[
            \sum_{i:\alpha_i<y}\nu_i
            =\frac{z_\ell}{M}
            \le F(\alpha_\ell)\le F(y).
      \]
      Thus~\eqref{pub-main-cdf-condition} holds. Lemma~\ref{pub-main-work}
      gives the stated rounded-work bound.
\end{proof}

\subsection{Exact Sums and Finite-Machine Inequalities}

\begin{proof}[Proof of Lemma~\ref{pub-main-certificate}]
      The rational construction gives the exact coefficients
      \begin{align*}
            \mu+\frac\beta Z
              & = \frac{17061993668440506814801440321355077007}{6087762161272353248853235028830797824}, \\
            V & = \frac{45041789022865043205136711}{16071023398612989725000000}.
      \end{align*}
      Direct rational comparisons show that
      \[
            \frac{11}{4}<V<\mu+\frac\beta Z<\frac{2802671}{10^6}.
      \]
      The other quantities needed for feasibility satisfy the rational
      enclosures
      \begin{gather*}
            0.6428409359<Z<0.6428409360,\\
            1.1978096052<\mu<1.1978096053,\\
            0.1504386142<\delta_3<0.1504386143,
      \end{gather*}
      and the exact cap extrema are
      \[
            r_{\min}=\frac{272727273}{10^8},\qquad
            r_{\max}=\frac{298960373}{10^8}.
      \]
      These enclosures imply $3(r_{\max}-1)\delta_3<\mu$. Moreover,
      $r-r/(r-1)$ is increasing for $r>1$, so
      \[
            \frac{4(r_{\max}-1)}{3r_{\max}-2}
            <\frac87<\frac{240}{209}
            \le r_{\min}-\frac{r_{\min}}{r_{\min}-1}.
      \]
      Normalisation, the cumulative-weight condition, and domination
      follow from Lemma~\ref{pub-main-grid-validity} and
      \eqref{pub-main-domination}. The cap construction also gives
      $2<r_i<3$ and $\alpha_i+k_i\le r_i$. Hence all hypotheses of
      Lemma~\ref{pub-main-lift} hold.

      The companion verifier computes these finite sums and comparisons
      exactly. For rational $u\ge0$, it obtains $\lfloor\sqrt u\rfloor$
      by an integer square root. Likewise, the cap condition
      $r\ge r(\alpha_i)$ is equivalent to
      \begin{gather*}
            r\ge\frac{30}{11},\qquad r\ge\frac{\alpha_i+2}{2},\\
            r^2-(\alpha_i+2)r+\alpha_i\ge0,\qquad
            r^2\ge\xi+\frac{\xi-\alpha_i}{\eta}.
      \end{gather*}
      Thus neither the square roots nor the displayed decimal enclosures
      introduce a floating-point assumption.
\end{proof}

\end{document}